\documentclass[review]{elsarticle}

\usepackage{hyperref}
\usepackage{amsmath,amssymb}
\usepackage[utf8]{inputenc}

\journal{Optik – International Journal for Light and Electron Optics}

\begin{document}

\begin{frontmatter}

\title{Porosity and roughness determination of  porous silicon thin films  by genetic algorithms}

\author[a,b]{C. F. Ramirez-Gutierrez\corref{mycorrespondingauthor}}
\cortext[mycorrespondingauthor]{Corresponding author}
\ead{cframirezg@comunidad.unam.mx}

\author[c]{J. D. Casta\~no-Yepes}

\author[d]{M. E. Rodriguez-Garcia}

\address[a]{Posgrado en Ciencia e Ingenier\'ia de Materiales, Centro de F\'isica Aplicada y Tecnolog\'ia Avanzada, Universidad Nacional Aut\'onoma de M\'exico Campus Juriquilla, C.P. 76230, Qro., M\'exico}
\address[b] {Ingenier\'ia F\'isica, Facultad de Ingenier\'ia, Universidad Aut\'onoma de Quer\'etaro, C.P. 76010 Quer\'etaro, Qro., Mexico}
 \address[c]{Instituto de Ciencias Nucleares, Universidad Nacional Aut\'onoma de M\'exico, M\'exico Distrito Federal, C. P. 04510, M\'exico}
 
\address[d] {Departamento de Nanotecnolog\'ia, Centro de F\'isica Aplicada y Tecnolog\'ia Avanzada, Universidad Nacional Aut\'onoma de M\'exico  Campus Juriquilla, C.P. 76230, Qro., M\'exico}

\begin{abstract}
The problem of determining the porous silicon (PSi) optical constants, thickness, porosity, and surface quality using just reflectance data is board employing evolutionary algorithms. The reflectance measurements were carried  out  of PSi films over crystalline silicon (c-Si) substrate, and the fitting procedure was done by using a genetic algorithm. The PSi is treated as a mixture  of c-Si and air. Therefore, its effective optical constants can be correlated with the porosity trough effective medium approximation (EMA).  The results show that  genetic fitting has a good  match  with the experimental measurements (Near UV-Vis reflectance) and the thickness obtained by scanning  electron microscopy (SEM). 
\end{abstract}


\end{frontmatter}


\section{Introduction}
\label{Intro}
Porous silicon (PSi) is a nanostructured material~\citep{Pfeiffer} obtained by electrochemical etching of crystalline silicon (c-Si). In the case of p-type c-Si,  microporous can be obtained ($<2$ nm) if the resistivity of the sample is more than 0.1 $\Omega$ cm,  mesoporous ( $2-50$ nm) for heavy doped c-Si with resistivity between $0.1-0.001$ $\Omega$ cm~\citep{SailorBook}. For n-type c-Si with 0.1-0.01 $\Omega$ cm mesoporous are formed, while for more resistive n-Si macroporous formation ($>50$ nm) is present~\citep{SailorBook}.
Therefore, it is possible to tune the PSi electrical and optical properties through the porosity by using a combination of fabrication parameters such as the current density,  electrolyte composition, temperature~\cite{SailorBook, CFRamirez2017,  CFRamirez2016}, thermal oxidation\citep{Pfeiffer}, among others extrinsic and extrinsic parameters. \\
 PSi has a high surface area, diverse surface chemistry~\citep{Sailor1}, porous morphology~\citep{ZhangES},  high-efficiency photo- and electro- luminescence~\citep{CFRamirez2018, Torres-CostaRev}, as well as piezoelectric \citep{ARAGON2016177} and piezooptic properties~\citep{Vinikman}. Furthermore, PSi has effective optical properties that depend on the material that can fill the pores. The characteristics mentioned above make the PSi an interesting material for chemical sensing~\cite{Pavesi}, biological applications~\citep{Weiss},  photonics, and optoelectronics~\citep{Weiss}.\\
 It is common to use ex-situ techniques such as scanning electron microscopy (SEM), atomic force microscopy (AFM)~\citep{Pfeiffer}, profilometry, and gravimetry~\citep{Foss} to characterize  PSi properties. In some cases, the probe can be  destructive.  Other works are focused on describing the kinetics of the chemical reaction and following  the formation of the porous film in real time to determine  some properties of PSi in-situ~\citep{CFRamirez2017, CFRamirez2016, CFRamirez2018, Foss, RaoFTIR,  Andrusenko2012}. For any application of the PSi,   a completed knowledge of PSi optical properties is required, interface quality, thickness, and porosity. Optical transmittance and reflectance are good options because they are non-destructive techniques and provide rapid and accurate information of optical properties of c-Si and PSi in the visible range~\citep{BIRGIN}.\\
The calculation of the optical constants (refractive index $n(\lambda)$, and extinction coefficient, $k(\lambda)$) is usually made by fitting the reflectance or transmittance spectra. However, the methods are not trivial, in fact, they represent an inverse problem. This problem of estimating optical constants and  thickness using only transmittance or reflectance data has been addressed by using fitting procedures and optimization algorithms.  However, in the case of PSi, due to the nature of the random porous formation, it is necessary to consider other parameters such as porosity and interface roughness because this variety of inhomogeneities causes light scattering.  These inhomogeneities and roughness are no longer negligible,  and they can introduce a significant error in the determination of the optical constants~\citep{Guo}.

 Optimization algorithms solve this kind of problems, especially evolutionary algorithms because they can avoid local minima following many search pats simultaneously~\cite{Torres-Acosta2004, Bumroongsri2012, BIRGIN2003109, CFRamirez2014}. Torres-Acosta et al.~\citep{Torres-Acosta2004}  used a self-adaptive genetic algorithm to determine optical constants and thickness of  PSi films in the visible range (400-800 nm). However,  the fitting procedure used a parametrization of the real part of the refractive index and it did not consider the porosity and roughness of the  PSi film.  Nevertheless, it is possible to use the same methodology to introduce the porosity and interface roughness,  fit the reflectance spectrum and to determine the optical constants.\\
 To add the porosity percentage and roughness interface, the electrical permittivity of PSi can be described as an effective medium.  This method takes into account the system as a mixture composed by a host medium with $\hat{\epsilon}_m$ (c-Si) with inclusions within it, characterized by   $\hat{\epsilon}_i$, which allows the determination of the optical properties in the linear regimen. Thus, the PSi is modeled as an effective medium~\citep{ChoyBook, Huet} that is the result of a  mixture of c-Si and the material that fill the pores that can be a gas or  liquid.
 
To overcome this problem, this work proposes a methodology based on genetic algorithms to fit the near-specular reflectance (6$^\circ$  incidence) spectrum of single films of PSi over the c-Si substrate. A simultaneous determination of PSi properties such as optical constants, thickness,  porosity, and interface roughness is made only by using reflectance measures.  The system is considered as  a silicon (c-Si) single crystal, and by using an effective medium approximation (EMA)  the algorithm can determine the  refractive index $\eta(\lambda)$,  extinction coefficient $\kappa(\lambda)$, PSi thickness,  interfaces RMS roughness ($\sigma$) (Air/PSi and  PSi/Si substrate)  and the porosity. The model is tested by using several films of p-type Si fabricated with different anodization times.  Also, the thickness of the PSi samples was determined by the genetic fit and compared with the thickness obtained by electron scanning microscopy (SEM) images. 


 \section{Experimental section}
  \subsection{Porous silicon fabrication}
 Four  PSi films were fabricated by electrochemical etching using hydrofluoric acid (HF) in aqueous media.  Heavy boron doped Si (p$^{++}$) with 0.005 $\Omega$ cm of resistivity and $[100]$ crystalline orientation it was used. The samples were cut into squares and cleaned by the RCA standard method. The etching was carried out by using an electrolyte composed of HF/ ethanol in 3:7 volume ratio and a regimen of a constant current of 20 mA/cm$^2$. The porous formation was followed by photoacoustic using the methodology proposed by Ramirez-Gutierrez et al.~\citep{CFRamirez2018, CFRamirez2017, CFRamirez2016}.
Each sample was fabricated under the same conditions (current density, temperature, and electrolyte composition). The etching time was the changed parameter. Sample S1 was etched for 28 s,   S2 for 56 s,  S3 for 84 s, and  S4 for 112 s. 
 \subsection{Near specular reflectance }
Optical characterization of PSi films was carried out  using a Perkin Elmer UV-Vis Spectrophotometer Lambda 35 in the near-normal (6$^\circ$) relative specular reflectance mode from 1100 to 210 nm range. The spectrophotometer was self-calibrated using an aluminum mirror, and the samples were measured over the Si substrate. It means that the reflectance is the optical response of the PSi thin film  Si substrate (PSi/Si structure). \\
The spectra were corrected following the Eqs.~\ref{Eq1} and~\ref{Absolute}   to obtain the absolute reflectance.

\begin{equation}
R(\lambda)=R_{relative}(\lambda)R_{transfer}(\lambda),
\label{Eq1}
\end{equation}
\begin{equation}
R_{transfer}(\lambda)=\frac{R_{Teo}^{(Si)}(\lambda)}{R_{relative}^{(Si)}(\lambda)},
\label{Absolute}
\end{equation}

 where $R(\lambda)$ is the absolute reflectance of the sample (PSi/Si- structure),  $R_{transfer}$ is the transfer function of the spectrophotometer  built  trough theoretical reflectance  of Si ($R_{Teo}^{(Si)}$) and  the measured reflectance of a Si substrate ($R_{relative}^{(Si)}$).
\subsection{Morphological studies}
 \begin{figure}
\centering
\includegraphics[scale=.44]{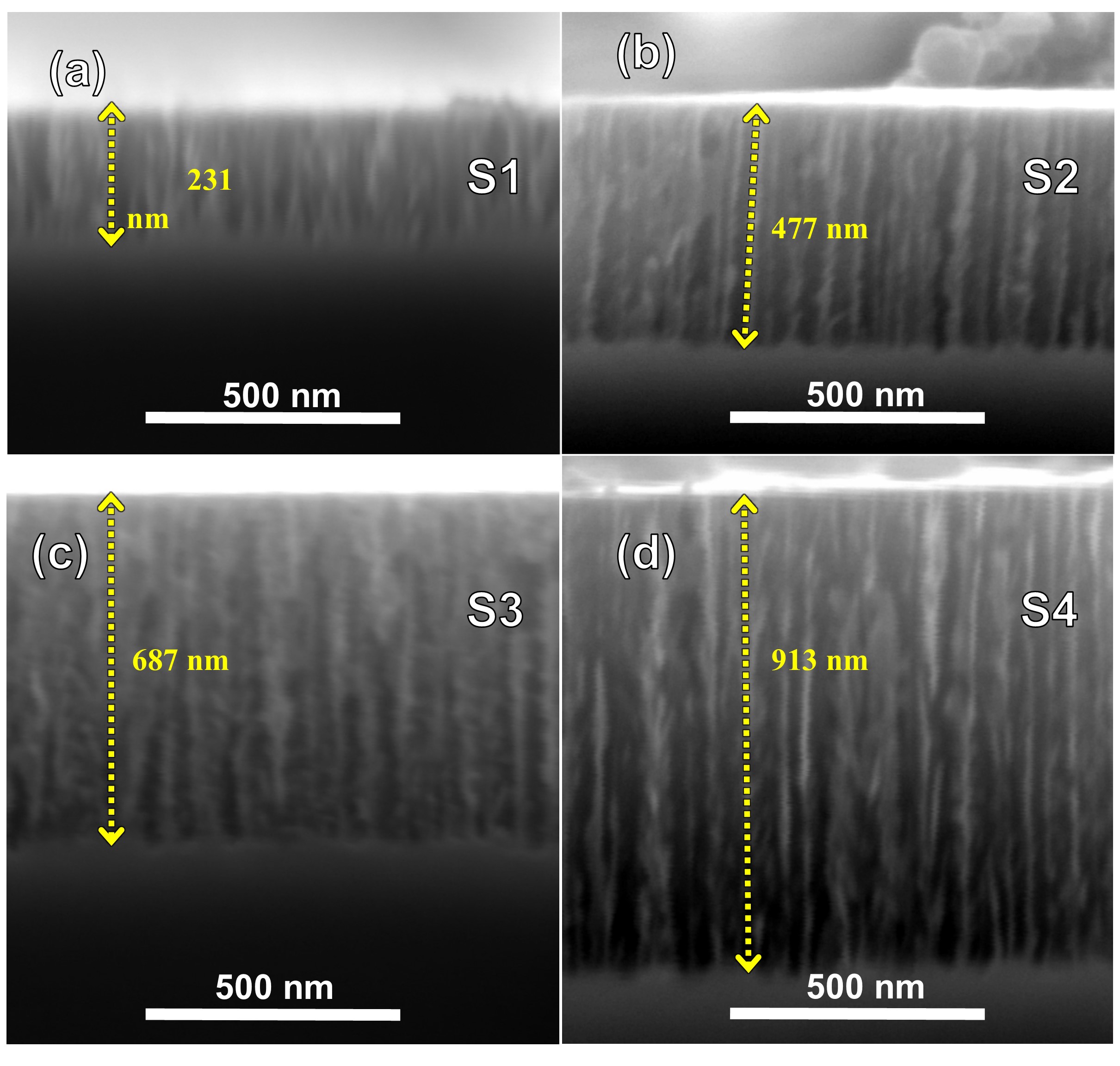}
\caption{Cross section of PSi samples obtained at different etching time.}
\label{SEMcross}
\end{figure}
After the optical characterization,  the samples were cut to determine the films thickness.  A MIRA3 TESCAN microscopy was used with a 5.0 kV electron acceleration voltage. Before the analysis, samples were fixed on the holder with copper tape. The samples were not covered with gold before the SEM analysis. Fig.~\ref{SEMcross} shows the SEM images of the cross-section of each sample. The ImageJ software was used for the image analysis and the determination of the thickness. Besides,  on the Fig.~\ref{SEMcross}   it is appreciable that the porous has not flat termination and exhibits lateral branching. 
\section{Thin-film calculations}
The method of the optical  admittance  is valid in the case of a linear regimen, that is very convenient because the optical constants can be expressed regarding dielectric constant of the medium $(\hat{\varepsilon}=\varepsilon_1+i\varepsilon_2)$ as follows:
\begin{eqnarray}
\textup{Re}(\hat{N})=\eta=\sqrt{\frac{\left |\hat\varepsilon  \right |+\varepsilon_1  }{2}},\nonumber\\
\textup{Im}(\hat{N})=\kappa=\sqrt{\frac{\left |\hat\varepsilon  \right |-\varepsilon_1  }{2}}.
\label{OpticalConstants}
\end{eqnarray}
The Eqs.~\ref{OpticalConstants} are useful because the dielectric constant of two-component materials (host and filling material) can be expressed through the effective medium approximation (EMA)\citep{ChoyBook, LLL}, where the volume fraction of the filling material is directly related with the porosity. For these simulations, it was proved the Maxwell-Garnett, Bruggeman, and Looyenga EMA formulas. However, due to the porous morphology  is  recommended to use Looyenga EMA formula (Eq.~\ref{LLLF})~\citep{LLL}, because it does not considered a regular geometry of the incrustations.
\begin{eqnarray}
\hat{\epsilon}_{eff}^{\frac{1}{3}}=(1-p)\hat{\epsilon}_{Si}^{\frac{1}{3}}+p\hat{\epsilon}_{air}^{\frac{1}{3}}
\label{LLLF}
\end{eqnarray}

\begin{figure}
\centering
\includegraphics[scale=0.4]{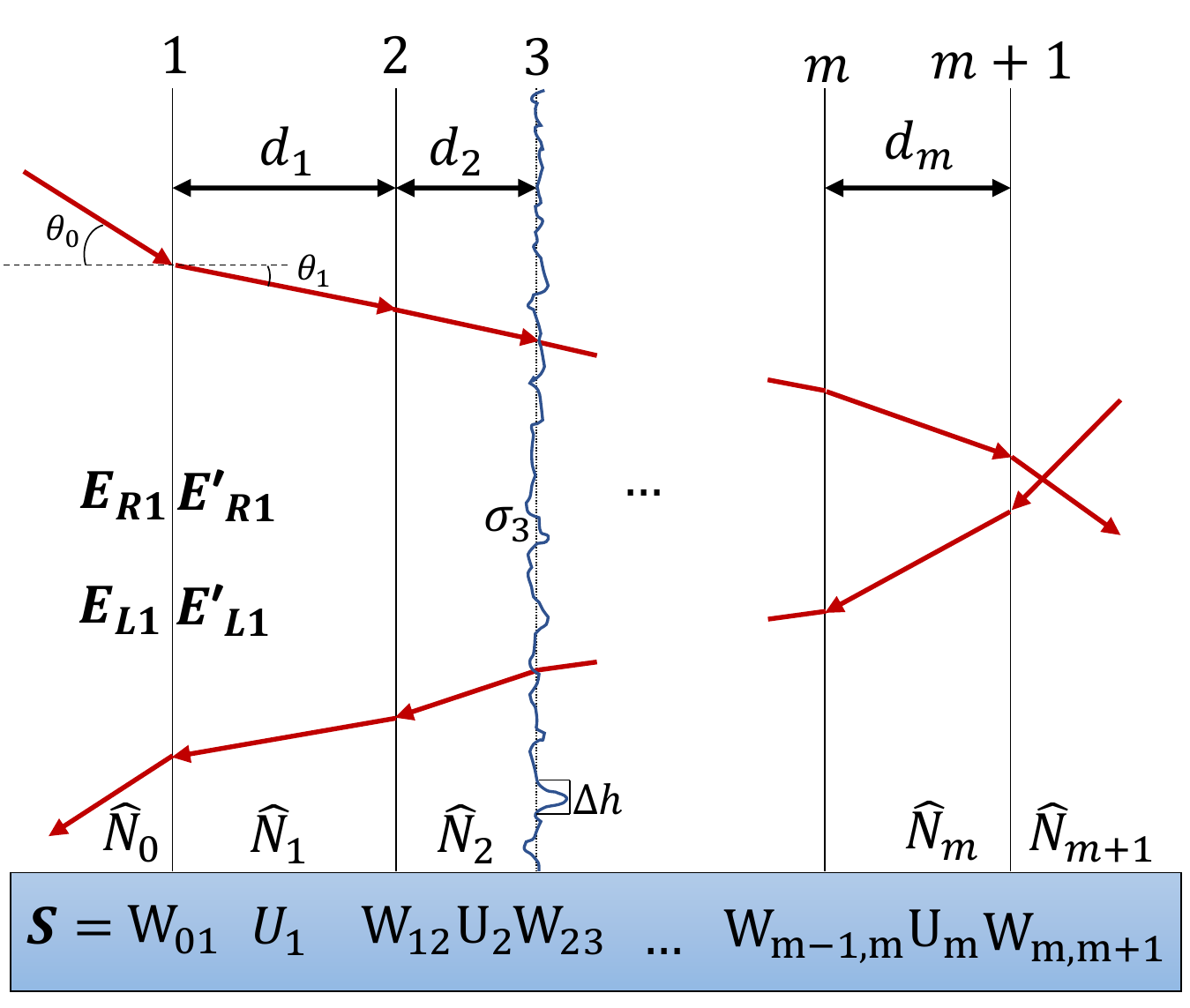}
\caption{Multilayer structure form with $m+1$ interfaces. This scheme shows the field amplitudes of the moving waves from left to right. The parameter $\sigma_3$ represents the RMS roughness of the third interface and $\Delta h$ the size of the irregularities in nanometers. It is assumed that $\Delta h<<\lambda$.}
\label{Propagations}
\end{figure}

To determine the reflectance  of multilayers systems, the optical admittance method~\citep{Mitsas}  introduces the transfer matrix $\mathbf{S}$ of the complete system through the multiplication of the refractive matrix $\mathbf{W}_{i-1,i}$ of each interface  and phase matrix $\mathbf{U}_{i}$ of each single film, where their components are expressed regarding the Fresnel coefficients. This notation indicates that $i$-th interface is the $i$-th material  to the right of the interface and it is numbered as is shown in Fig.~\ref{Propagations}.
\begin{eqnarray}
\mathbf{W}_{i-1,i}=\frac{c_{i-1,i}}{t_{Ri}}\left ( \begin{matrix}
1 & -r_{Li}\\r_{Ri} 
 & t_{Ri}t_{Li}-r_{Ri}r_{Li}
\end{matrix} \right ),
\label{AdMatrix}
\end{eqnarray}
where ${r_{Ri,Li}}^{(0)}$ and ${t_{Ri,Li}}^{(0)}$ are the usual Fresnel coefficients defined for optical admittance~\citep{Mitsas} of the $i$-th interface. The coefficients written in the form showed in the Eq.~\ref{FresnellR} introduce the RMS roughness $(\sigma_i)$ of the $i$-th interface.
\begin{eqnarray}
r_{Ri}&=&r_{Ri}^{(0)}\textup{exp} \left [ -2(2\pi\sigma_i\hat{N}_{i-1}/\lambda)^2 \right ]=\alpha r_{Ri}^{(0)},\nonumber\\
r_{Li}&=&r_{Li}^{(0)}\textup{exp} \left [ -2(2\pi\sigma_i\hat{N}_{i}/\lambda)^2 \right ]=\beta r_{Li}^{(0)}, \nonumber \\
t_{Ri}&=&t_{Ri}^{(0)}\textup{exp} \left [ -1/2(2\pi\sigma_i/\lambda)^2(\hat{N}_{i}-\hat{N}_{i-1})^2 \right ]=\gamma t_{Ri}^{(0)}, \nonumber\\
t_{Li}&=&t_{Li}^{(0)}\textup{exp} \left [ -1/2(2\pi\sigma_i/\lambda)^2(\hat{N}_{i-1}-\hat{N}_{i})^2 \right ]=\zeta t_{Li}^{(0)}.
\label{FresnellR}
\end{eqnarray}

The  $c_{i-1,i}$ parameter is related with the light polarization given by
\begin{eqnarray}
c_{i-1,i}=\left\{\begin{matrix}
\cos \theta_{i-1}/\cos \theta_{i} &\textup{ for p-polarization} \\ 
1 & \textup{ for s-polarization}
\end{matrix}.\right.
\label{C-Polarization}
\end{eqnarray}
The phase matrix   is  defined by Eq.~\ref{PhaseM}, where $\hat{N}_{i}$ is the complex refraction index and $d_i$  is  the medium thickness.
\begin{eqnarray}
\mathbf{U}_{i}=\begin{pmatrix}
\textup{exp} \left ( i \frac{2\pi}{\lambda} \hat{N}_{i} d_i \right ) & 0\\ 
0 & \textup{exp} \left (- i \frac{2 \pi}{\lambda} \hat{N}_{i}d_i \right )
\end{pmatrix}.
\label{PhaseM}
\end{eqnarray}
Finally, the transfer matrix of the  multilayer structure is defined as
\begin{eqnarray}
\mathbf{S}=\mathbf{W}_{01}\mathbf{U}_{1}\mathbf{W}_{12}\mathbf{U}_2...\mathbf{W}_{m,m+1}=\begin{pmatrix}
s_{11} &s_{12} \\ s_{21} &s_{22} 
\end{pmatrix}
\label{TransferMatrix},
\end{eqnarray}
and the reflectance and transmitance  are determined by
\begin{eqnarray}
R=\left | r_R \right |^2&=&\left | \frac{s_{21}}{s_{11}} \right |^2 ,\nonumber\\
T=\left | t_R \right |^2&=&\left | \frac{1}{s_{11}} \right |^2.
\label{RT}
\end{eqnarray}
This method permits to calculate the total response of multilayer systems  based on porous silicon (PSi) such as distributed Bragg  reflector (DBR)\citep{MANIYA2014828} or resonator cavities.\\

\section{Genetic fit}\label{GenSec}

The genetic algorithm used in this work is useful to determine the optical constants,  porosity,  roughness, and thickness of PSi by using  its reflectance spectrum and an effective medium approximation~\cite{ChoyBook, LLL}. In the case of absolute reflectance of thin films stacks, the optical response is dependent on several optical and structural parameters. Here, each parameter will be a gene, and the complete array of this genes is the chromosome (${\bf V}$) that is defined as follows 

\begin{equation}
 {\bf V}=(v_1, v_2,v_3,v_4)=\left(p,d,\sigma_{0},\sigma_{1}\right),
\label{Gen1}
\end{equation}
where $p$ is the porosity, $d$ the PSi thickness, $\sigma_{0}$ the roughness in the interface air/PSi, and $\sigma_{1}$  the roughness in the interface PSi/Si.
The parameters defined in the chromosome are used to calculate the theoretical reflectance spectrum $(R_G(\lambda))$, and it is compared with the experimental $(R_{exp}(\lambda))$ trough the  penalty function (Eq.~\ref{Fit1}). $R_{exp}(\lambda)$ is the absolute reflectance of PSi/Si structure obtained by using  Eq.~\ref{Eq1}.
Thus,  the method to estimate the parameters related in the chromosome (Eq.~\ref{Gen1}) is a problem of least squares fitted between as a measured and theoretical reflectance. 
\begin{equation}
F({\bf V})=\sum_{i=1}^{N}\left [ R_{exp}(\lambda_i)- R_{G}(\lambda_i, p, d,\sigma_0,\sigma_1)\right ]^2.
\label{Fit1}
\end{equation}
The values of some genes are constraints~\citep{Bumroongsri2012,BIRGIN2003109} in order to guarantee values with physical sense. Indeed, the initial population and the next generations have to satisfy the condition of the Eq.~\ref{Restric}.
\begin{eqnarray}
0<p<1\textup{   for all } \lambda \, \epsilon \left [ \lambda_{min},\lambda_{max} \right ],\nonumber\\
\sigma_i<100 \textup{  nm} \textup{   for all } \lambda \, \epsilon \left [ \lambda_{min},\lambda_{max} \right ],
\label{Restric}
\end{eqnarray}
The genetic algorithm used for this calculations is described in the next steps and it is based on the works of references~\citep{Torres-Acosta2004,Bumroongsri2012, BIRGIN2003109}.
\begin{enumerate}
\item \textbf{Population}: a number of $J_{P_0}$ individuals (chromosomes) are created by choosing a  random value for every single gen. The values of genes have to satisfy the constrain conditions  (Eq.~\ref{Restric}).
\item \textbf{Reproduction}:  each individual $V_{father}$ has given a number $K_{off}$ offspring. For that,  another individual of the same generation called $V_{mother}$ is select randomly.   The offspring is given by two reproduction ways determined with $\rho$ probability. 
 \begin{eqnarray}
v_i^{son}=\left \{ \begin{matrix}
v_i^{(father)}\textup{  if  } r<\rho_1 \\ 
v_i^{(mother)}\textup{  if  } r\geqslant \rho_1,
\end{matrix} \right.
\label{RPoly}
\end{eqnarray}
 \begin{eqnarray}
v_i^{son}=\frac{v_i^{(father)}+v_i^{(mother)}}{2},
\label{RMono}
\end{eqnarray}
where $\rho_1$ is the probability of inheriting the father's gene and $r$ is a random number used to decide it.\\
\item  \textbf{Mutation}: a certain number of individuals are mutated to introduce new genes in order to avoid local minima. The mutation is generated in the chromosome as following:
 \begin{eqnarray}
\tilde{v}_i^{son}=v_i^{son}\left(1+ \mathcal{N }_{(0,1)}\tau \right),
\label{Mutation}
\end{eqnarray}
 where $\tau$ is a weight function parameter and $\mathcal{N }_{(0,1)}$ is a random value from a normal distribution.
 \item \textbf{Family competition}: every  chromosome is used to simulated the reflectance spectrum by using the Looyenga EMA rule (Eq.~\ref{LLLF}) and the optical admittance method (Eqs.~\ref{TransferMatrix} and~\ref{RT}). The penalty function (Eq.~\ref{Fit1}) is evaluated for each individual. After,  the family competition starts. The individual who has the best fit is the best adapted, it  survives to the next generation and becomes a new father. If the new generation gives back a better fit, the mutation value $\tau$ is reduced by a factor $f$. Finally,  the steps mentioned above are repeated a $J_{G_0}$ times.  The value $J_{G_0}$  is the number of generations, and the individual with the lower value of penalty function represent the bets adapted and is the solution to the problem.
\end{enumerate}

\section{Results}
\begin{figure}
\centering
\includegraphics[scale=0.5]{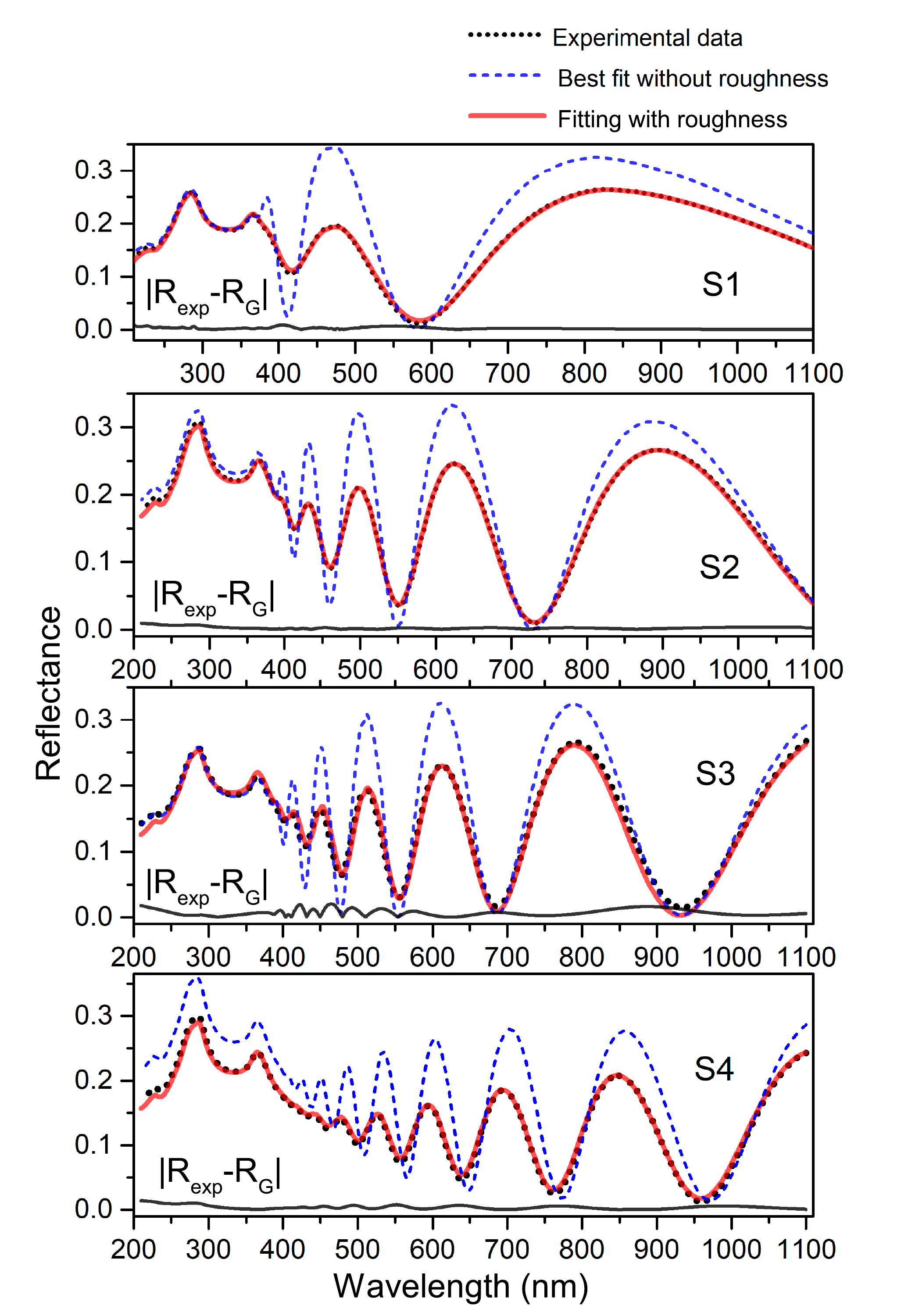}
\caption{Measured reflectance of PSi layers (black dash line), genetic fit without roughness (blue dash line), genetic fit with roughness (red line), and comparison between simulated and measured reflectance (black line)}
\label{ReflectFit}
\end{figure}

Using the genetic fit, some results from the analysis of measured reflection signals are presented. For all simulations, a $J_{P_0}=50$ individuals were used for each generation with $K_{off}=15$ offsprings  during $J_{G_0}=100$ generations.  The parameters obtained were the films thickness, porosity, interface roughness, and etch rate. Further, a comparison between thickness obtained by a genetic fit and by SEM is given. Fig.~\ref{ReflectFit} shows the experimental and the best fit of the reflectance spectrum for each sample, and table~\ref{Table1} shows the value of the fitting  parameters obtained.


In all samples, the current density during the anodization was $20$ mA/cm$^2$.  It means, due to the self-limited character of the  reaction~\citep{George, Keshavarzi} the porosity has to remain constant for short anodization times. This  fact can be noted in the value of the porosity in table~\ref{Table1}. As it was reported elsewhere,  the etch rate is almost constant \citep{SailorBook,CFRamirez2016}. Therefore, it is expected that the thickness of the samples S2, S3, and S4 are closely an integer  multiple value of the thickness of  sample S1.
As can be seen, the thickness obtained by genetic fit is close to the  obtained value by SEM. However, in all cases, the genetic fit reports  lower values than SEM. Even so, the obtained values for genetic fit are in the uncertainty range of the SEM technique, that for this case is $\pm10$ nm.\\
\begin{table}[]
\centering
\caption{Shows the parameter values of best fit for  UV-Vis spectra of  PSi films shown in the Fig.~\ref{ReflectFit} and the comparison between the thickness measured by SEM and by the genetic fit.}
\begin{tabular}{ c c  c  c  c c c}
\hline
Sample & $d$(nm)  &$d\pm10$ (nm)& Porosity&$\sigma_0$ (nm) & $\sigma_1$ (nm)& Etch rate \\ 
 & Genetic Fit &SEM&     & Air/PSi &PSi/Si& nm/s \\\hline
S1 & 225 &    231& 0.63 & 8.05 &24.88 &8.50 \\
S2   &467& 477  & 0.62         &    5.89       &22.90 &8.36 \\
S3   & 648&687   & 0.62&9.55   &23.79  &8.08\\
S3   & 900&913&0.62  &3.51           & 37.45&8.07\\
\hline
\label{Table1}
\end{tabular}
\end{table}
The main effect of the interface roughness  is the loss of reflected intensity due to the scattering.  Also, the branching of the porous can contribute to the scattering. This effect can be appreciated in Fig.\ref{ReflectFit} where the  genetic fit it was ran without roughness correction ( blue dash line). It is clear that the best fit   has always a higher intensity than the measured reflectance because the scattering is neglected.\\
As SEM images show (Fig.~\ref{SEMcross}),  the interface Air/PSi is smooth, and the values of the $\sigma_0$ obtained by genetic fit are lower than 10 nm. This means, that the major contribution of the scattering is the PSi/Si interface. It was attained for $\sigma_1$  values from 24.88 nm for S1 to 37.45 nm for S4. It is noticeable that $\sigma_1$ value increases monotonously as a function of the etching time. Nevertheless, this fact is not directly related  with a high roughness of PSi/Si interface. The increment of $\sigma_1$ is an indication of a thicker film, because there are more branched porous through the light path.\\
\begin{figure}
\centering
\includegraphics[scale=0.4]{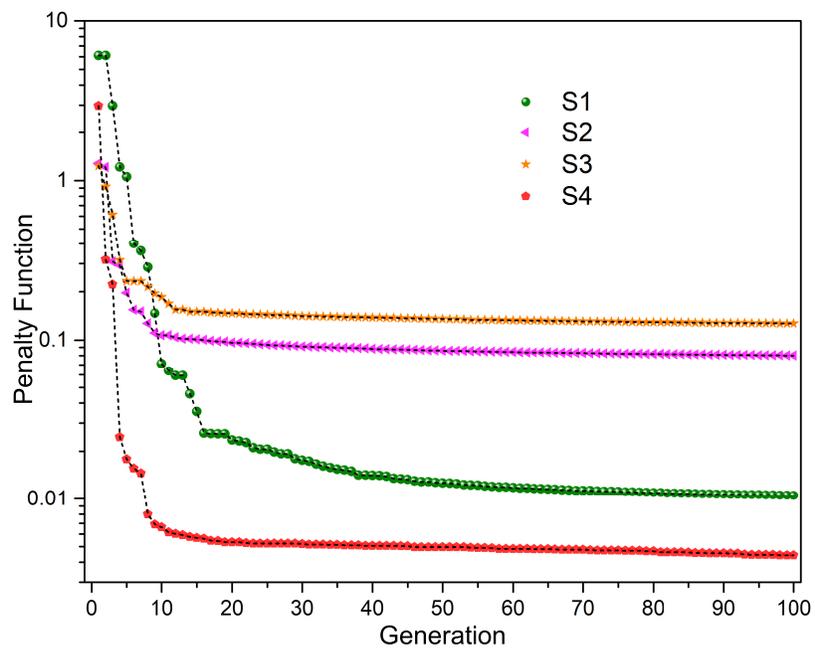}
\caption{Penalty function evaluation progress. Each point represents the best adapted individual of each generation.}
\label{PenaltyFit}
\end{figure}
Finally, Fig.~\ref{PenaltyFit} shows the evolution of the penalty function during 100 generations (Eq.~\ref{Fit1}). The behavior of the penalty function shows that the algorithm finds the solution during the first generations. After that, the algorithms stay in a local minimum even if the mutation parameter increases. This means that there is a bottom edge that the genetic fit cannot overcome. The fitting results of this methodology  is dependent on the quality of the measured spectrum and the EMA model used to simulate the theoretical reflectance.
\section{Conclusions}
By using the algorithm described in the section~\ref{GenSec},  the optical parameters of four PSi thin films  was determined. The algorithm determines the porosity, thickness, and interface roughness simultaneously. In this case, the anodization current  remained constant at 20 mA/cm$^2$ and the etching time was changed  in order to determine the etching rate and the evolution of the interface roughness.  It was found that under the etching conditions used for these experiments, the self-limited character of the reaction is kept, and the average porosity and etch rate are constants.\\
PSi has no stable surface because it has a diversified chemical surface (Si-H, Si-C, -Si-O, Si-N). During  PSi aging,  some surface modifications take place, such as silicon oxides formation. This methodology does not consider the contribution of another surface chemical species or the participation of the Si$_x$O$_y$ thin film that was formed during the passage of time.\\
The main problem related with the estimation of optical parameters of PSi films is connected with good UV-Vis experimental measurements. For this reason, it is always recommended to correct the reflectance measurements by using a  reflectance pattern, in this case, the reflectance spectra were corrected by using the theoretical reflectance of c-Si to obtain the absolute reflectance.
Finally, this methodology represents a powerful tool to determine the optical and morphological parameters of PSi thin films just by using the reflectance spectrum in the UV-Vis range that represents an advantage over microscopy techniques or ellipsometry. 
\section*{Acknowledgments}
C.F. Ramirez-Gutierrez and J.D. Casta\~no-Yepes want to thank Consejo Nacional de Ciencia y Tecnolog\'ia M\'exico (CONACYT), for the financial support of their Ph.D. studies. Thanks to Julian Ramirez-Gutierrez M.Sc. and  Beatriz Mill\'an-Malo Ph.D., for their assistance in the simulation. English editing by Kevin Steven Maya-Mart\'inez.

\end{document}